\documentclass{elsart}

\usepackage{graphicx,amsmath,amsfonts,amssymb}

\newcommand{\ket}[1]{\ensuremath{|{#1\rangle}}}

\renewcommand{\eqref}[1]{Eq.~(\ref{#1})}
\renewcommand{\cal}{\mathcal}
\newcommand\twofigs[2]{
\includegraphics[width=0.495\columnwidth]{#1}
\includegraphics[width=0.495\columnwidth]{#2}}

\begin{document}

\begin{frontmatter}



\title{Quantum Parrondo's games using quantum walks}


\author[ad1]{Adrian P. Flitney}
%\corauth[cor]{Corresponding author.}
\ead{aflitney@unimelb.edu.au}
\address[ad1]{School of Physics, University of Melbourne, VIC 3010, Australia}

\begin{abstract}
% Text of abstract
We study a quantum walk in one-dimension using two different ``coin'' operators.
By mixing two operators,
both of which give a biased walk with negative expectation value for the walker position,
it is possible to reverse the bias through interference effects.
This effect is analogous to that in Parrondo's games, where alternating two losing
(gambling) games can produce a winning game.
The walker bias is produced by introducing a phase factor into the coin operator,
with two different phase factors giving games $A$ and $B$.
We give the range of phases for which the Parrondo effect can be obtained
with $A$ and $B$ played alternately
or in other (repeated) deterministic sequences.
The effect is transitory.
For sufficiently large times the original bias resumes.
\end{abstract}

\end{frontmatter}

% main text
%\section{}
%\label{}

\maketitle

\section{Introduction}
%copied from one of my reviews
In Parrondo's game~\cite{harmer99b}, in its usually form as a gambling game,
a single agent plays against a bank, with the choice of two games $A$ and $B$,
whose results are determined by the toss of biased coins.
Each of these games is losing when played in isolation
but when played alternately or in some other deterministic or random sequence
(such as $ABAB$ etc.)
can become a winning game.
Because of its anti-intuitive nature, Parrondo's games are also referred to as Parrondo's paradox.
Game $B$ is constructed in such a way that there is a form of feedback from the game state that,
when the game is played by itself,
causes the game to be a losing game.
Interspersing plays of game $A$ disrupts the feedback and the net result can be a winning game.
The feedback can be in the form of dependence on the total capital of the player~\cite{harmer99a},
on the history of previous wins and losses~\cite{parrondo00,kay02}
or, in a multiplayer setting, on the performance of neighbouring players~\cite{toral02}.

Interest in quantum models of Parrondo's games was initiated by Meyer~\cite{meyer02a,meyer02b},
who presented a model analogous to the capital-dependent Parrondo's game,
and was later extended to the history-dependent model~\cite{flitney02b,flitney04b}.
A close quantum analogue of the capital dependent Parrondo's game
was also developed by Ko\u{s}\'{i}k~\cite{kosik07}.

A quantum analogue of a classical discrete random walk
can be obtained by replacing
the classical coin by an SU(2) operator.
If the walker position on a one-dimensional line represents the player's total capital,
a quantum walk becomes a natural setting for a quantum Parrondo's game.
As distinct from the classical case,
in a quantum setting the change from winning to losing
may be achieved by interference effects between the two games.

In the quantum Parrondo's game model of Ref.~\cite{chand11}
two players A and B play alternately.
Winning for player A is defined as motion to the right
while motion to the left is defined as winning for player B.
It is then obvious that when
A and B use their coins alternately that one or the other of them will `win',
or that the expectation of the walker position will be zero,
in which case the players are joint winners, according to the authors' criteria.
This does not have the anti-intuitive aspect that is characteristic of Parrondo's
games nor is it a single player with a choice of two games.
A two-dimensional analogue of this idea has also appeared as an e-print~\cite{ampadu11}.
Despite its failure to meet the criterion as a Parrondo's game the idea
used in Refs.~\cite{chand11,ampadu11}
of biasing a quantum walk using phase factors
can be used to construct a version of a quantum Parrondo's game that relies on quantum interference
to produce the counter-intuitive motion of the walker in a one-dimensional quantum walk.

In this paper
we first introduce quantum walks
and discuss how to control them using the coin parameters and the initial state.
This is well covered in the literature
%(for reviews of quantum walks see \cite{kempe03a,va12})
and is included here briefly only for completeness.
In Sec.~\ref{sec:our_model} the new quantum Parrondo's model is presented
and Sec.~\ref{sec:results} presents results for all combinations of games $A$ and $B$ of length four or less,
giving the regions in parameter space for which the effect can be achieved.
It will be noticed that although the expectation of the walker position
is reversed for some finite number of time steps,
eventually the original bias reasserts itself.
Thus we term this phenomena {\em transient} quantum Parrondo's games.

\section{Basics of quantum walks}
\label{sec:qwalks}
The usefulness of classical random walks in mathematics, theoretical computer science
and computational physics~\cite{papa94,motwani95}
has inspired significant interest in quantum walks, 
both in continuous-time~\cite{farhi98,childs02}
and in discrete-time~\cite{aharonov93,meyer96,aharonov01,ambainis01}.
%An additional degree of freedom
%or quantum ``coin'',
%is required to achieve unitarity in discrete-time, discrete-space, quantum walks~\cite{meyer96}.
Quantum walks reveal a number of startling differences to their
classical counterparts.
In particular, diffusion on a line is quadratically faster \cite{nayak00,travaglione02},
while propagation across some graphs is exponentially faster \cite{childs02,childs03}.
Quantum walks also show promise as a means of implementing some quantum algorithms~\cite{shenvi03,ambainis04}.
An overview of quantum walks is given in Ref.~\cite{kempe03a}
and a recent comprehensive review appears in Ref.~\cite{va12}.

If a quantum particle moving in discrete time steps along a line
is updated at each step, in superposition, to the left and right,
the global process is necessarily non-unitary~\cite{meyer96}.
However, an additional degree of freedom,
the chirality, taking values $L$ and $R$,
allows unitary quantum walks to be constructed.
Let ${\cal H}_{\rm P}$ be the Hilbert space of discrete particle positions in one-dimension,
spanned by the basis $\{|x\rangle : x \in \mathbb{Z}\}$.
%In each time-step the particle will move according to the chirality of the quantum coin,
%to the left or right.
Let ${\cal H}_{\rm C}$ be the Hilbert space of chirality, or ``coin'' states,
spanned by the orthonormal basis $\{|L\rangle, |R\rangle\}$.
A discrete-time, discrete-space quantum walk
on the Hilbert space ${\cal H}_{\rm P} \otimes {\cal H}_{\rm C}$
consists of a unitary operation $\hat{U}$ on the coin state,
followed by the updating of the position to the left or right
according to the result, that is,
\begin{equation}
\hat{E} = (\hat{S} \otimes \hat{\cal P}_{R} \;+\; \hat{S}^{-1} \otimes \hat{\cal P}_{L})
        (\hat{I}_{\rm P} \otimes \hat{U}),
\end{equation}
where $\hat{S}$ is the shift operator in position space,
$\hat{S} |x\rangle = |x+1\rangle$,
$\hat{I}_{\rm P}$ is the identity operator in position space,
and $\hat{\cal P}_R$ and $\hat{\cal P}_L$ are projection operators on the coin space
with $\hat{\cal P}_R + \hat{\cal P}_L = \hat{I}_{\rm C}$, the coin identity operator.
For example, a walk controlled by an unbiased quantum coin
can be carried out by the transformations
\begin{equation}
\label{eq:step1}
\begin{split}
|x, L \rangle &\rightarrow \, \frac{1}{\sqrt{2}}
        \left( |x-1, L \rangle \,+\, i |x+1, R \rangle \right), \\
|x, R \rangle &\rightarrow \, \frac{1}{\sqrt{2}}
        \left(i |x-1, L \rangle \,+\, |x+1, R \rangle \right).
\end{split}
\end{equation}
To achieve an unbiased quantum walk with Eq.~(\ref{eq:step1})
we require an initial state
$|\psi_0\rangle = (|0, L \rangle - |0, R \rangle)/\sqrt{2}~$.
The scheme is then equivalent to the Hadamard quantum walk
with initial state $(|0, L\rangle + i |0, R\rangle)/\sqrt{2}$.

Tregenna {\em et al.}~\cite{tregenna03} has shown that with a general initial state
\begin{equation}
\ket{\psi(x,0)} =
	\left( \sqrt{\eta} \ket{R} + e^{i \mu} \sqrt{1-\eta} \ket{L} \right) \otimes \ket{0},
\end{equation}
and with a coin operator of maximum generality (up to an overall phase)
\begin{equation}
\label{eq:general_flip}
\hat{U}(\rho, \theta, \phi) = \begin{pmatrix}
											\sqrt{\rho} 				 & \;e^{i \theta} \sqrt{1-\rho} \\
											e^{i \phi} \sqrt{1-\rho} & \;-e^{i (\theta + \phi)} \sqrt{\rho}
			 						   \end{pmatrix},
\end{equation}
$\phi$ disappears in the final dynamics,
while $\theta$ only appears in the combination $\theta + \mu$.
Both $\rho$ and $\eta$ have non-trivial effects on the dynamics
but an unbiased walk requires $\rho = \frac{1}{2}$.

\section{Parrondo's game constructed from alternating coin phases}
\label{sec:our_model}
In the following we will chose the initial state
$|\psi_0\rangle = (|0, L \rangle - |0, R \rangle)/\sqrt{2}~$,
and a more symmetrical form of the coin operator,
\begin{equation}
\label{eq:flip}
\hat{U}(\rho, \alpha) = \begin{pmatrix} e^{i \alpha} \sqrt{\rho} & i \sqrt{1-\rho} \\
                         i \sqrt{1-\rho} & e^{-i \alpha} \sqrt{\rho}
                \end{pmatrix},
\end{equation}
obtained by multiplying the operator in Eq.~(\ref{eq:general_flip})
by an overall phase $e^{i \alpha}$,
and setting $\theta = \phi = \frac{\pi}{2} - \alpha$.
With $\rho = \frac{1}{2}$ and $\alpha=0$ an unbiased quantum walk is obtained.
However, with $\rho = \frac{1}{2}$ a bias can be introduced,
to the left or right,
by the use of a non-zero phase factor $\alpha$.
If we restrict $\alpha \in [-\frac{\pi}{2}, \frac{\pi}{2}]$,
then $\alpha > 0$ causes average motion to the left while $\alpha < 0$ causes motion to the right.
Introduce the operators
\begin{equation}
\hat{A}(\alpha) = \hat{U}(\frac{1}{2}, \alpha), \;
\hat{B}(\alpha) = \hat{U}(\frac{1}{2}, \beta),
\end{equation}
for some $\alpha, \beta \in [\frac{-\pi}{2}, \frac{\pi}{2}]$.
Connection with Parrondo's games can now be obtained by choosing to use $\hat{A}$ and $\hat{B}$
for games $A$ and $B$, respectively,
with both $\alpha, \beta > 0$.
Thus, quantum walks generated by either $\hat{A}$ or $\hat{B}$ alone
are biased towards the negative (`losing').
The loss is an approximately linear function of the number of time steps $t$.
For example,
$\langle x \rangle |_{t=50} \, \simeq -15.0 \, \sin \alpha$
and $\langle x \rangle |_{t=100} \, \simeq -29.6 \, \sin \alpha$.
%as indicated in Fig.~\ref{f:A100}.
However, for some choices of $\alpha$ and $\beta$,
and some combinations of $A$ and $B$,
interference effects can reverse the dynamics
giving a positive expectation value of the walker position (`winning'),
at least up to some finite number of time steps.
Alternately playing $A$ then $B$ $n$-times from the initial state $|\psi_i \rangle$
produces a final state $|\psi_f \rangle$ given by
\begin{equation}
|\psi_f \rangle = ( \hat{B} \otimes \hat{A} )^n \, |\psi_i \rangle.
\end{equation}
The final states for other sequences are produced analogously.
%The following section gives results for these sequences in detail.

%\begin{figure}
%\begin{center}
%\includegraphics{A1000.eps}
%\end{center}
%\caption{The expectation value of the walker position after 1000 time steps
%for game $A$
%for a range of values of the phase parameter $\alpha$ [color online].}
%\label{f:A1000}
%\end{figure}

\section{Results}
\label{sec:results}
\begin{figure}
\begin{center}
\includegraphics{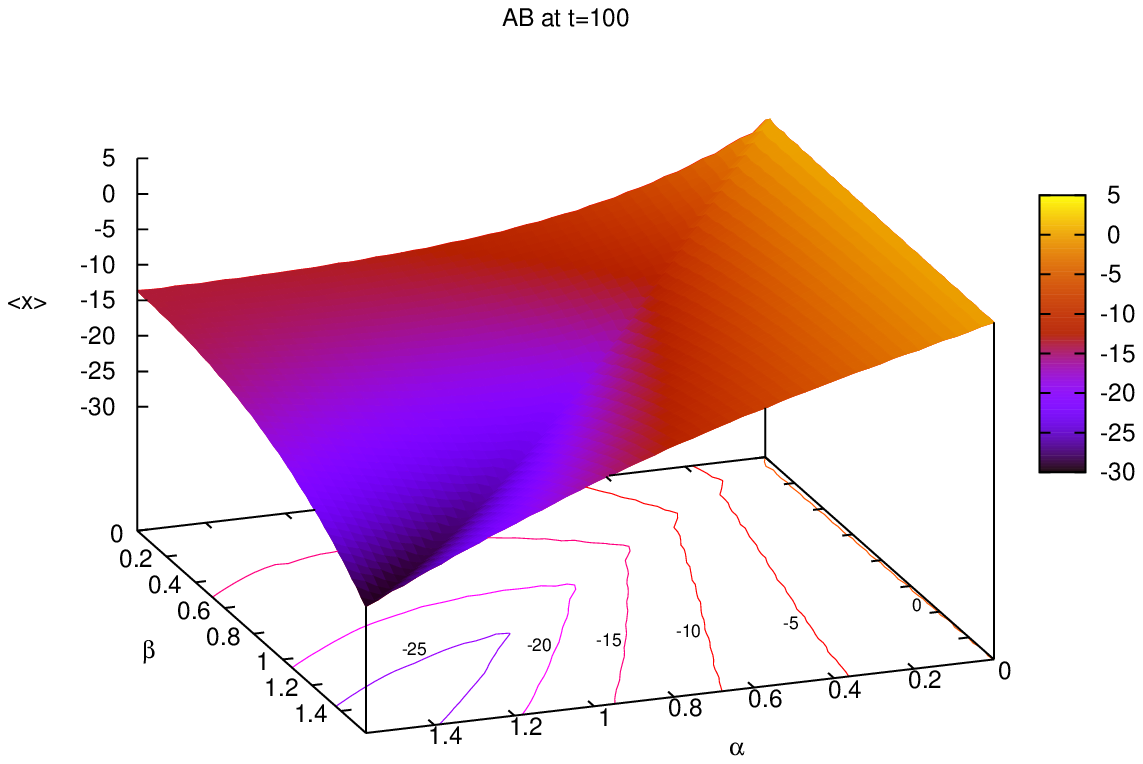}
\end{center}
\caption{The expectation value of the walker position after 100 time steps
(50 repetitions of the sequence)
for an alternating sequence of games $A$ and $B$
for a range of values of the phase parameters $\alpha$ and $\beta$;
set $\alpha = \beta$ for game $A$ alone [color online].}
\label{f:AB100}
\end{figure}

\begin{figure}
\begin{center}
\includegraphics{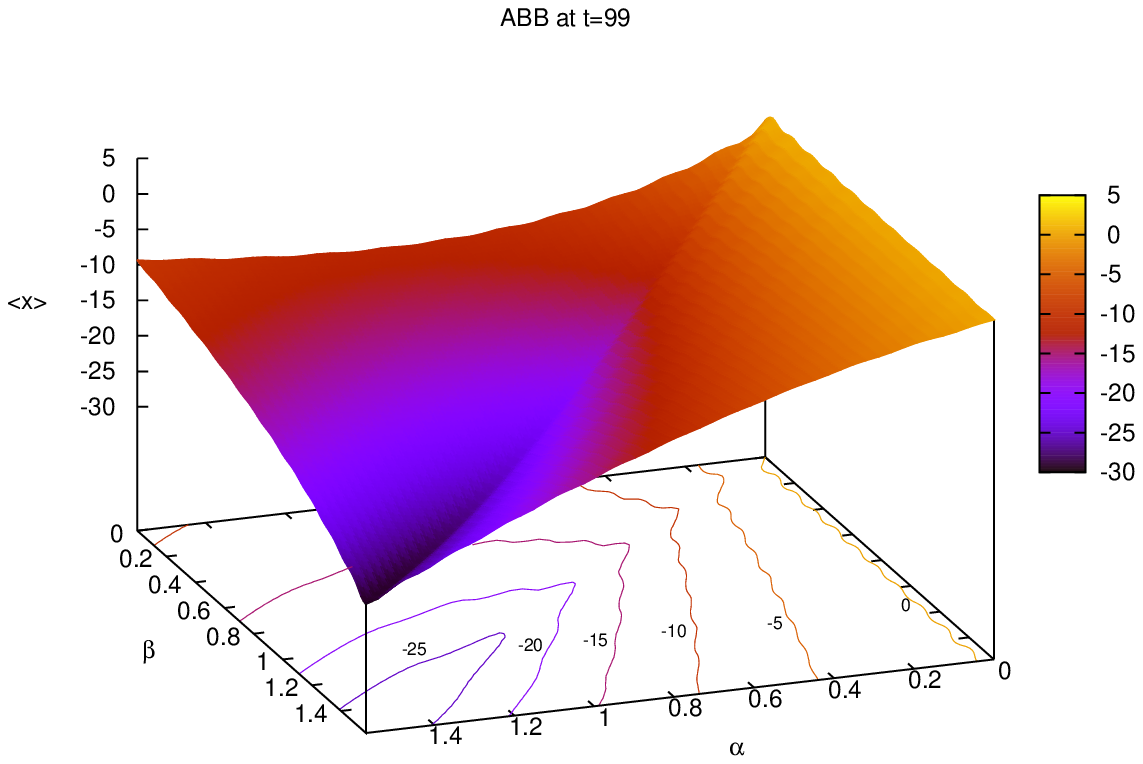}
\end{center}
\caption{The expectation value of the walker position after 99 time steps
(33 repetitions of the sequence)
for the sequence of games $ABB$,
for a range of values of the phase parameters $\alpha$ and $\beta$;
set $\alpha = \beta$ for game $A$ alone [color online].}
\label{f:ABB99}
\end{figure}

\begin{figure}
\begin{center}
\includegraphics{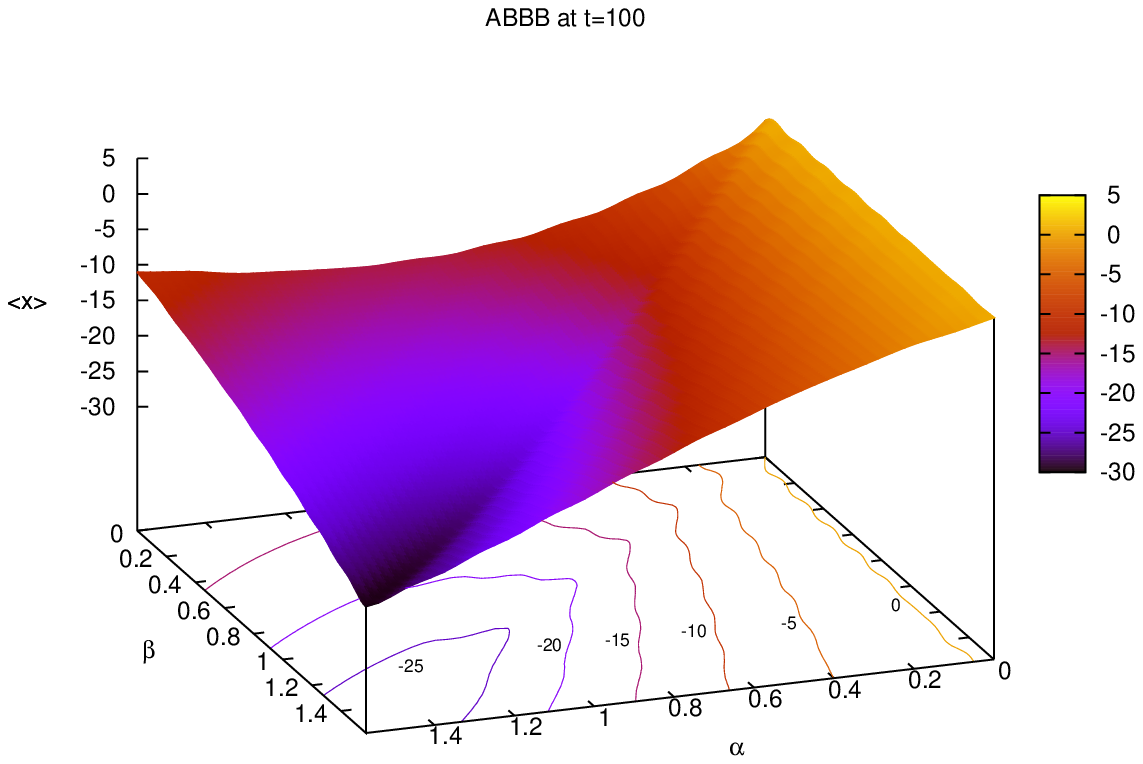}
\end{center}
\caption{The expectation value of the walker position after 100 time steps
(25 repetitions of the sequence)
for the sequence of games $ABBB$,
for a range of values of the phase parameters $\alpha$ and $\beta$;
set $\alpha = \beta$ for game $A$ alone [color online].}
\label{f:ABBB100}
\end{figure}

All repeated sequences of four games were tested.
Of these, only $AB$, $ABB$ and $ABBB$ were found to give some positive expectation for the walker
position after 100 time steps.
In each case there is only a small region of values of the phase parameters $\alpha$ and $\beta$
that give rise to the positive positions,
as indicated in Figs~\ref{f:AB100}--\ref{f:ABBB100}.
Note that interchanging games $A$ and $B$ only has the effect of reversing the role of $\alpha$ and $\beta$
and gives no new dynamics.
The results for game $A$ alone can be obtained by setting $\alpha = \beta$.
In each case this results in a more negative value of $\langle x \rangle$
for a given number of time steps $t$
than the alternating sequence for near-by values of $\alpha \ne \beta$.

\begin{figure}
\begin{center}
\twofigs{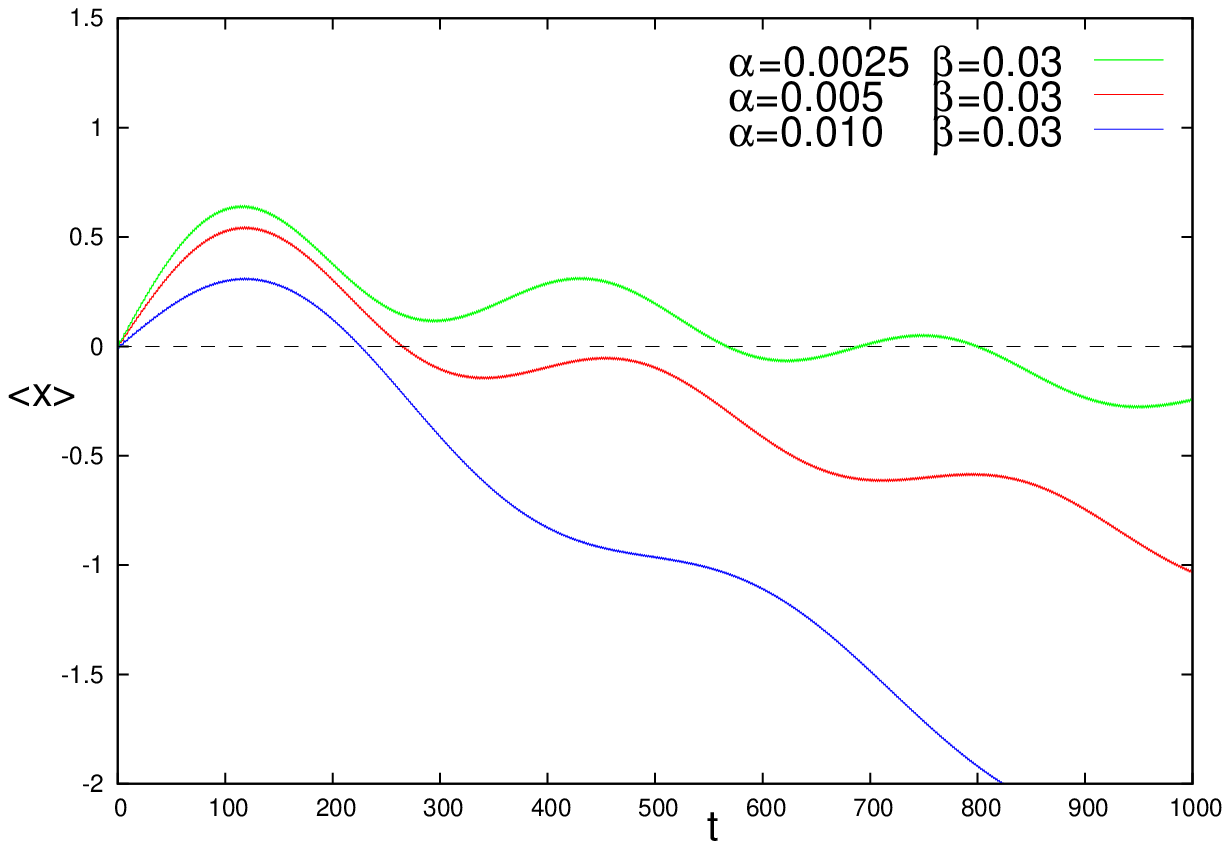}{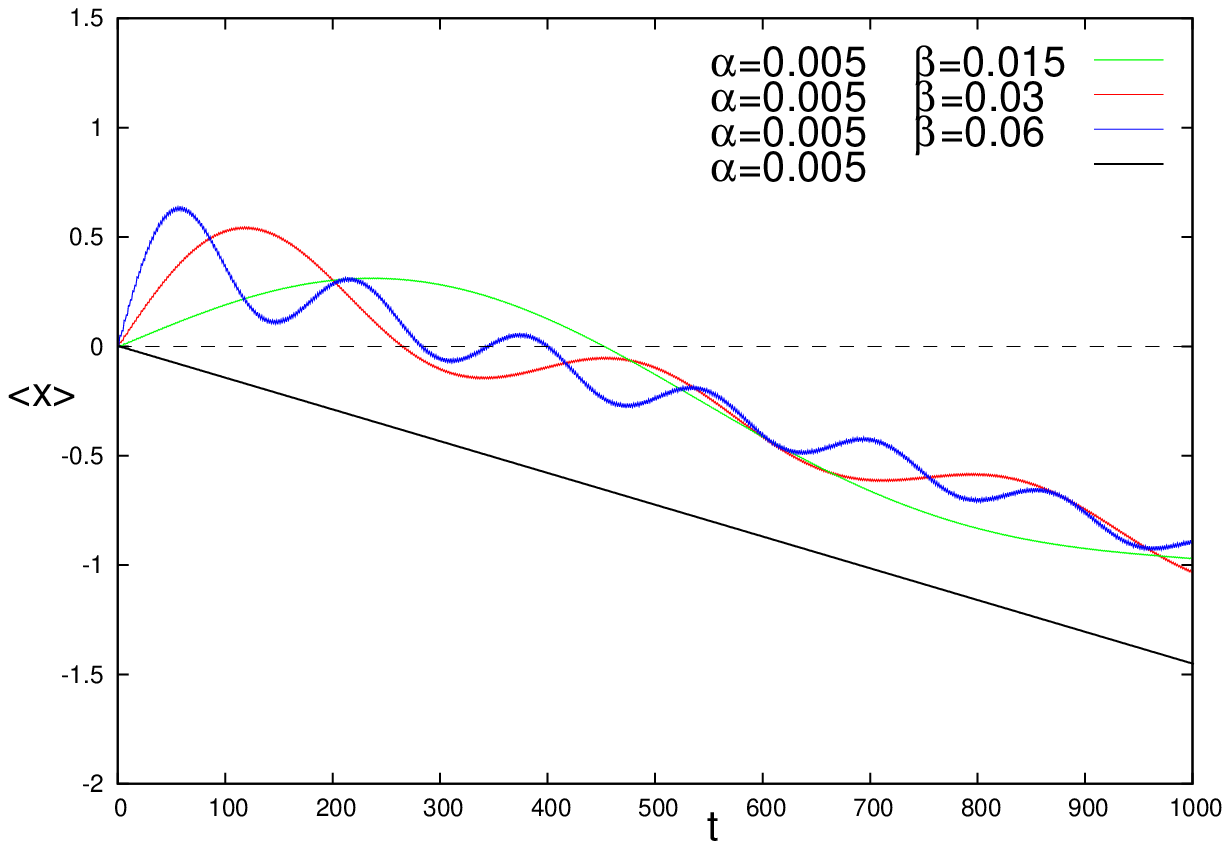}
\end{center}
\caption{The expectation value of the walker position after 1000 time steps
for an alternating sequence of games $A$ and $B$
for (left) $\beta = 0.03$ and several values of $\alpha$
and (right) for $\alpha = 0.005$ and several values of $\beta$;
straight line is for game $A$ only [color online]. }
\label{f:AB_change_ab}
\end{figure}

\begin{figure}
\begin{center}
\twofigs{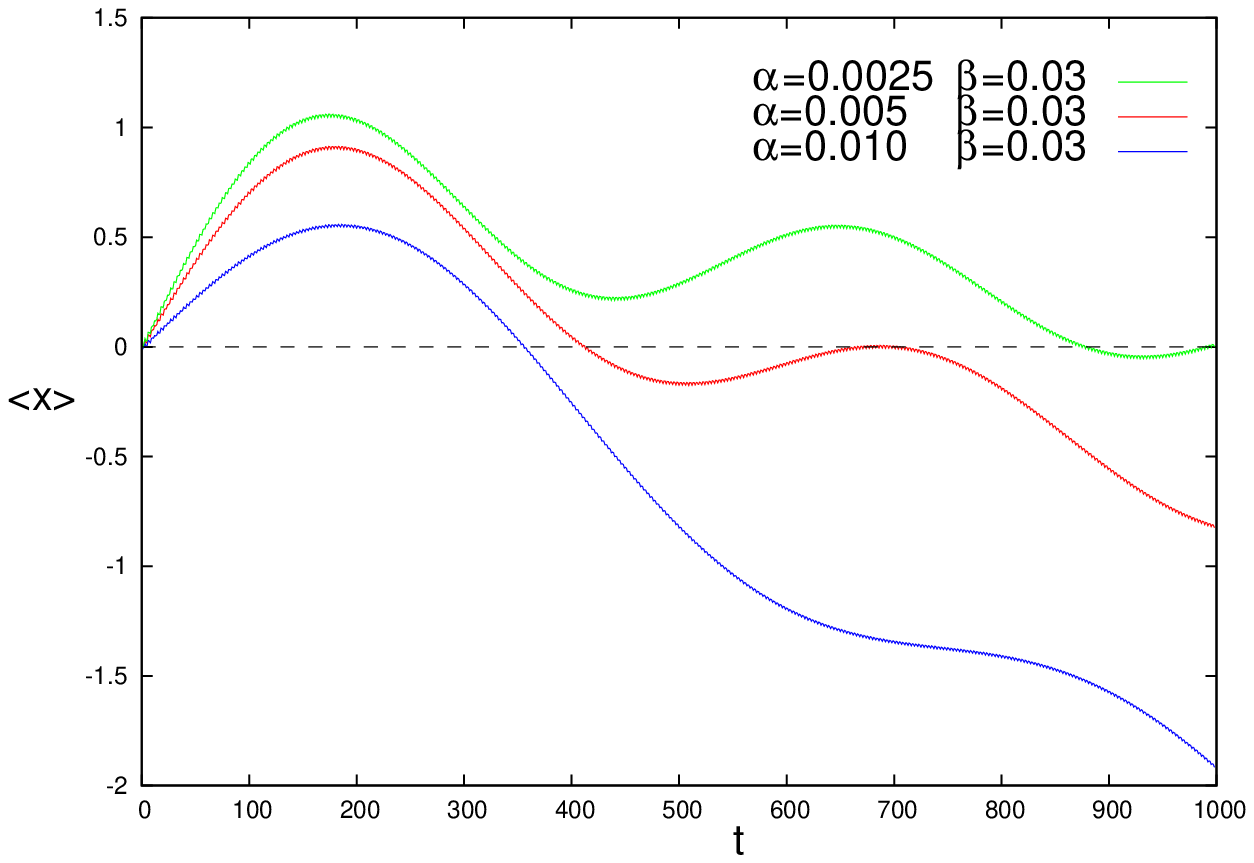}{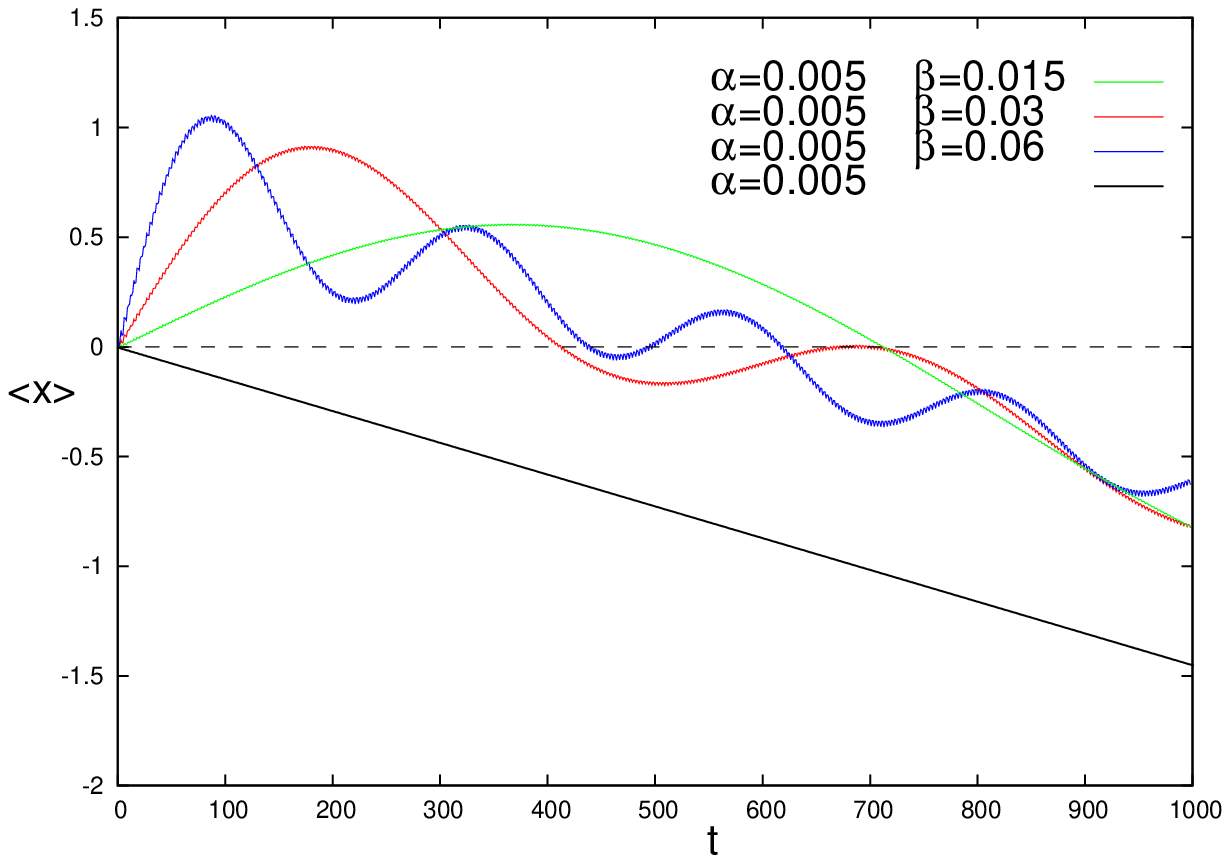}
\end{center}
\caption{The expectation value of the walker position after 1000 time steps
for the sequence of games $ABB$
for (left) $\beta = 0.03$ and several values of $\alpha$
and (right) for $\alpha = 0.005$ and several values of $\beta$;
straight line is for game $A$ only [color online]. }
\label{f:ABB_change_ab}
\end{figure}

\begin{figure}

\begin{center}
\twofigs{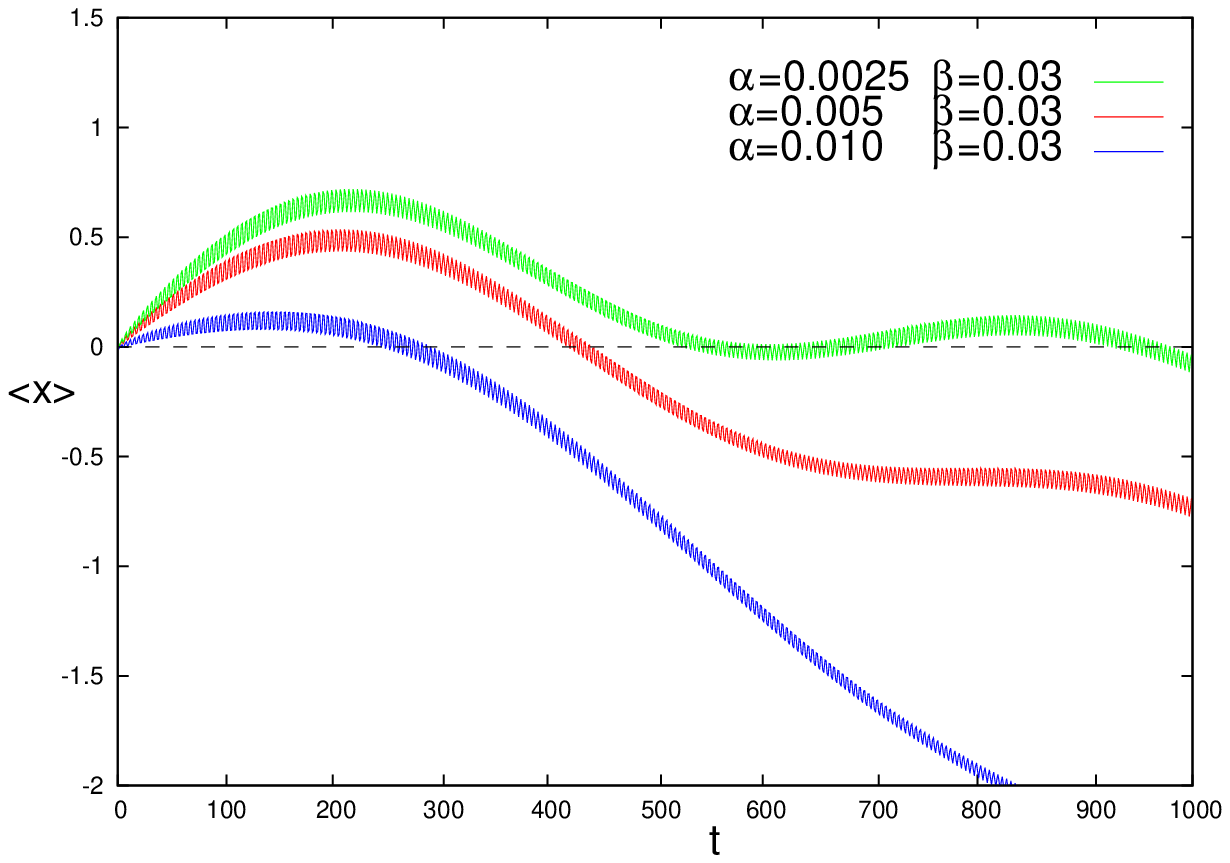}{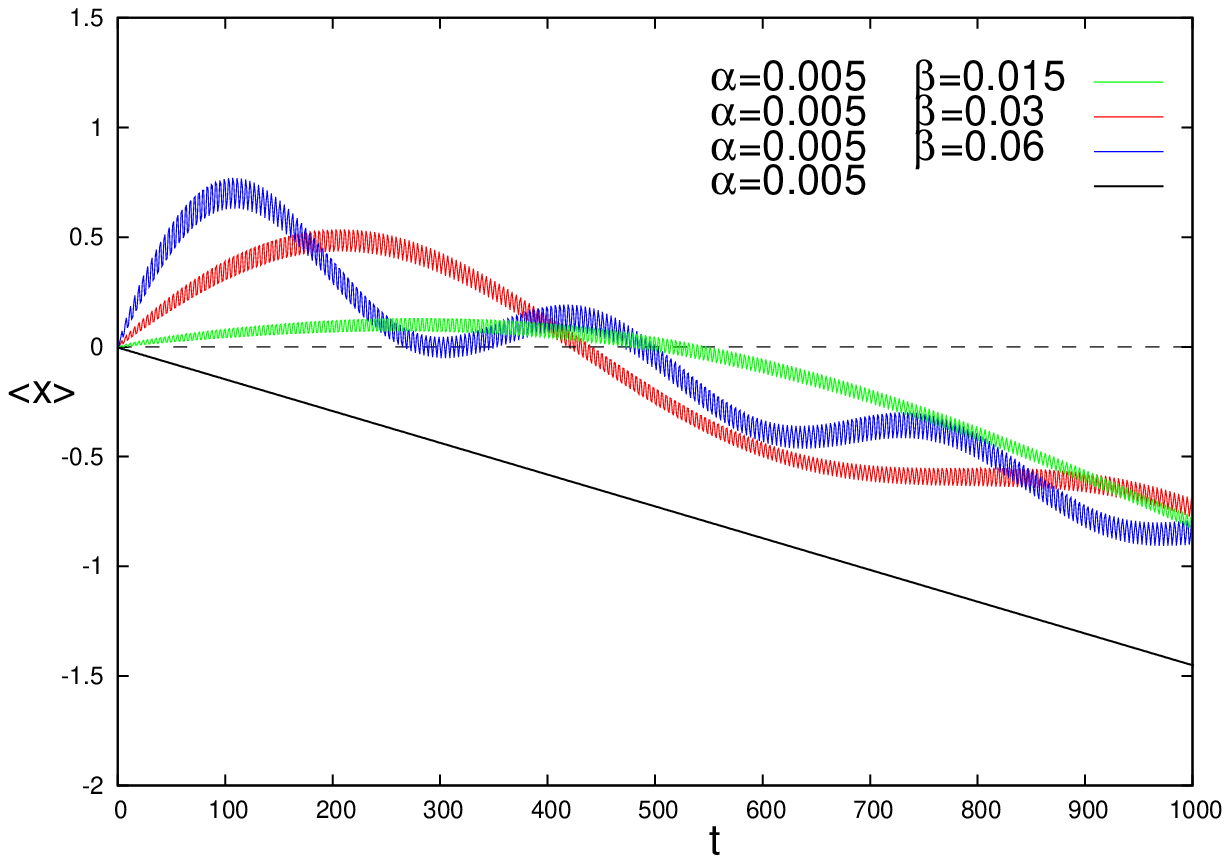}
\end{center}
\caption{The expectation value of the walker position after 1000 time steps
for the sequence of games $ABBB$
for (left) $\beta = 0.03$ and several values of $\alpha$
and (right) for $\alpha = 0.005$ and several values of $\beta$;
straight line is for game $A$ only.
The apparent thickness of the curves is due to low amplitude, short period oscillations.
These can be seen online with sufficient enlargement [color online]. }
\label{f:ABBB_change_ab}
\end{figure}

For a sufficiently large number of time steps
all sequences eventually show a return to a negative expectation value for the walker position.
%For this reason we have termed the phenomenon a {\em transient} quantum Parrondo's game.
Some examples of the expectation value of the walker for the first 1000 time steps are shown in
Figs.~\ref{f:AB_change_ab}--\ref{f:ABBB_change_ab},
for the sequences $AB$, $ABB$ and $ABBB$, respectively.
The solid (near) straight lines in the right hand figures indicate the result for playing game $A$ only,
for $\alpha = 0.005$.
These graphs indicate that $\langle x \rangle$ has an oscillatory pattern superimposed upon a downward trend.
Decreasing $\alpha$ leads to a lessening of the downward slope,
while decreasing $\beta$ leads to longer periods of oscillation.
A short period, low amplitude oscillation is superimposed upon the longer one.
The amplitude of this oscillation becomes more pronounced as we go from the sequence $AB$ to $ABB$ to $ABBB$,
as close inspection of the figures show.

\section{Conclusion}
We have shown that a quantum walk on a line can give rise to a transient quantum Parrondo's game.
Two different biased coin operators $\hat{A}$ and $\hat{B}$
play the role of the two games in the Parrondo effect.
Bias is achieved by adding a phase factor to the coin operator.
Although each operator,
when used alone,
gives negative expectation values for the walker position as the number of steps increases,
when used in the repeated sequences $AB$, $ABB$ or $ABBB$
can give rise to movement in the positive direction,
for some values of the phase factors.
We have detailed the range of values of these phase factors that give rise
to the Parrondo effect after 100 time steps
for each of these sequences.
The effect is transitory.
For large enough times all sequences eventually result in a negative
expectation value for the  walker position.

%\section*{Acknowledgments}
%\noindent

\vfill
\pagebreak



\end{document}